\documentclass[conference]{IEEEtran}
\IEEEoverridecommandlockouts

\usepackage{cite}
\usepackage{amsmath,amssymb,amsfonts}
\usepackage{graphicx}
\usepackage{textcomp}
\usepackage{xcolor}
\usepackage{url}

\makeatletter
\def\ps@IEEEtitlepagestyle{%
  \def\@oddfoot{\parbox{\textwidth}{\footnotesize\centering
    \copyright~2026 IEEE. Personal use of this material is permitted.
    Permission from IEEE must be obtained for all other uses, in any current
    or future media, including reprinting/republishing this material for
    advertising or promotional purposes, creating new collective works, for
    resale or redistribution to servers or lists, or reuse of any copyrighted
    component of this work in other works.}}%
  \def\@evenfoot{}%
}
\makeatother

\begin{document}

\title{WIP: LLM Odyssey: A Game-Based Platform for Teaching LLM Engineering Concepts}

\author{\IEEEauthorblockN{Priyamvada Tripathi}
\IEEEauthorblockA{\textit{Tufts Institute for Artificial Intelligence} \\
\textit{Tufts University}\\
Medford, MA, USA \\
pia.tripathi@tufts.edu}
}

\maketitle

\begin{abstract}
This work-in-progress (WIP) innovative practice category paper presents LLM Odyssey, an open source, browser-based serious gaming platform comprising 13 interactive games for teaching Large Language Model (LLM) engineering concepts. Topics such as tokenization, transformer architecture, prompt engineering, retrieval augmented generation (RAG), and production deployment are underrepresented in computer science curricula. Existing interactive tools address individual concepts but lack pedagogical scaffolding or structured learning pathways. LLM Odyssey addresses this gap through three learning tiers aligned with Bloom's revised taxonomy: Cognitive Core (7 foundational games), Systems Forge (5 production engineering games), and Foundry Arena (capstone challenges). Each game incorporates five pedagogical strategies drawn from the literature: immediate formative feedback, scaffolded hints grounded in the Zone of Proximal Development, progressive difficulty informed by flow theory, worked examples to manage cognitive load, and authentic scenarios drawn from production practice. The platform was deployed in Winter 2026 semester at a Canadian college for an initial review. Feedback confirmed functional requirements and identified adaptive difficulty as a priority for future development. A formal mixed methods evaluation protocol (N=50) has been designed, comprising pre and post knowledge tests, validated surveys, engagement analytics, and interviews, and is documented here to enable future evaluation studies with the publicly available platform.
\end{abstract}


\section{Introduction}
Demand for graduates with practical LLM engineering skills has grown substantially as Large Language Models have become widely deployed in software systems \cite{bommasani2022}\cite{huyen2022}. Industry practitioners identify proficiency in prompt engineering, retrieval augmented generation, fine tuning, and production deployment as necessary competencies that current CS programs do not consistently develop \cite{bommasani2022}\cite{huyen2022}. Students may complete degree programs with theoretical machine learning foundations but limited hands on exposure to the operational challenges specific to LLM systems. This gap motivates the need for dedicated instructional resources in LLM engineering.

LLM engineering presents instructional challenges that are distinct from general machine learning education. Prompt engineering requires iterative experimentation, spanning strategies from direct instructions to structured reasoning prompts such as chain-of-thought \cite{wei2022}, and has been shown to be difficult even for technically experienced practitioners without specific LLM background \cite{zamfirescu2023}. Teaching prompt engineering is itself a recognized emerging area of pedagogy \cite{denny2023}. Additional challenges include managing probabilistic outputs, context window constraints, and production concerns such as latency, cost, and safety.

No existing platform provides structured, comprehensive coverage of LLM engineering from foundational concepts through production practice. Static lectures on attention mechanisms can support recall without developing operational understanding. Textbook examples of prompt engineering do not reliably transfer to novel tasks. Production considerations such as API cost, failure modes, and compliance are largely absent from academic coursework. Tools such as OpenAI Playground and Hugging Face Spaces address individual concepts but lack progressive difficulty, pedagogical scaffolding, or structured learning pathways. This gap between available instructional resources and required competencies provides the direct motivation for LLM Odyssey.

Game based learning (GBL) has an established research base as an approach to technical skill acquisition. Hamari et al. \cite{hamari2014} conducted a systematic review of gamification studies and found generally positive, though context dependent, evidence of engagement effects. Black and Wiliam \cite{black1998} document that immediate formative feedback during practice supports learning. Brown, Collins, and Duguid \cite{brown1989} show that situating learning in authentic contexts supports knowledge transfer. Brennan and Resnick \cite{brennan2012} argue that interactive creation and reflection support computational thinking development. Taken together, these findings provide a literature based rationale for a GBL approach to LLM engineering education. Direct empirical evidence in this specific domain does not yet exist, and validation of this approach remains an open question for future studies.

This paper reports: (1) background and related work; (2) the platform design and architecture; (3) implementation challenges and preliminary expert feedback; and (4) the designed evaluation protocol. Section II reviews background and related work. Section III describes the platform design. Section IV reports implementation and preliminary findings. Section V concludes with future directions.

\section{Background and Related Work}
LLM engineering presents instructional challenges that general machine learning pedagogy does not fully address. Attention mechanisms, embedding spaces, and gradient flows are not directly observable; parameter interactions are frequently nonlinear and counterintuitive; and production concerns such as API cost, latency budgets, and failure modes are largely absent from academic coursework. These characteristics suggest that hands on, feedback rich environments may be better suited to developing LLM competency than lecture or textbook instruction alone.

Game based learning (GBL) offers a theoretically grounded response to these challenges. Hamari et al. \cite{hamari2014} conducted a systematic review and found generally positive, though context dependent, evidence of engagement effects from gamification. Black and Wiliam \cite{black1998} document that immediate formative feedback during practice supports learning. Brown, Collins, and Duguid \cite{brown1989} show that situating learning in authentic contexts supports knowledge transfer. Brennan and Resnick \cite{brennan2012} argue that interactive creation and reflection support computational thinking development.

The platform's design draws selectively on four established learning theories. Constructivism \cite{piaget1954} motivates the emphasis on active experimentation over passive reading. Vygotsky's scaffolding principle \cite{vygotsky1978} motivates graduated support calibrated to the learner's current capability. Mastery learning \cite{bloom1968} motivates unlimited retries with a 70\% completion threshold, prioritizing depth over speed. Self Determination Theory \cite{deci2008} motivates progress visualization and tier unlocks as supports for autonomy and competence, without reliance on external grade linked rewards.

No existing platform provides structured, progressive coverage of LLM engineering from foundational concepts through production practice. Tools such as OpenAI Playground and Hugging Face Spaces address individual concepts but lack pedagogical scaffolding or structured learning pathways. LLM Odyssey is designed to fill this gap. Whether GBL achieves the anticipated benefits in this specific domain is the central empirical question that future evaluation will need to address.

\section{Platform Design and Architecture}
\begin{figure}[!t]
\centering
\includegraphics[width=0.9\columnwidth]{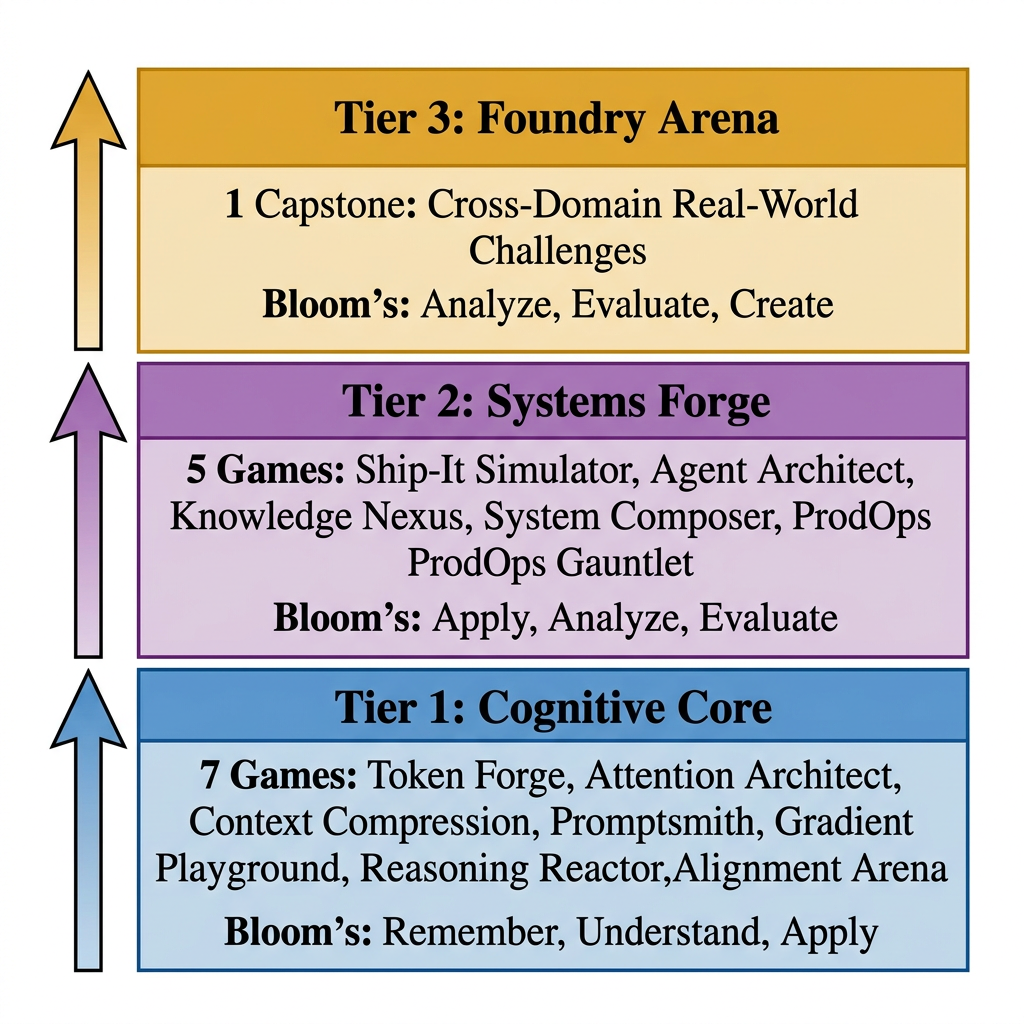}
\caption{LLM Odyssey contains three Tier Progression with 13 games.}
\label{fig:tiers}
\end{figure}

LLM Odyssey comprises 13 interactive games organized across three tiers that progress from foundational understanding to applied synthesis, mapped to Bloom's revised taxonomy (see Fig.~\ref{fig:tiers}) \cite{anderson2001}. Tier I, Cognitive Core (Games 1 to 7), targets the Remember, Understand, and Apply levels through parameter manipulation and real time observation of outcomes such as token counts, attention patterns, and loss curves. Tier II, Systems Forge (Games 8 to 12), targets the Apply, Analyze, and Evaluate levels through scenario based decisions under realistic production constraints including latency budgets, cost limits, and service level objectives. Tier III, Foundry Arena (Game 13), targets the Analyze, Evaluate, and Create levels through open ended cross domain challenges with embedded reflection prompts. This sequencing is intended to ensure foundational competency before production concepts are introduced.

\subsection{Game Mechanics and Pedagogical Design}
\begin{figure}[!t]
\centering
\includegraphics[width=0.9\columnwidth]{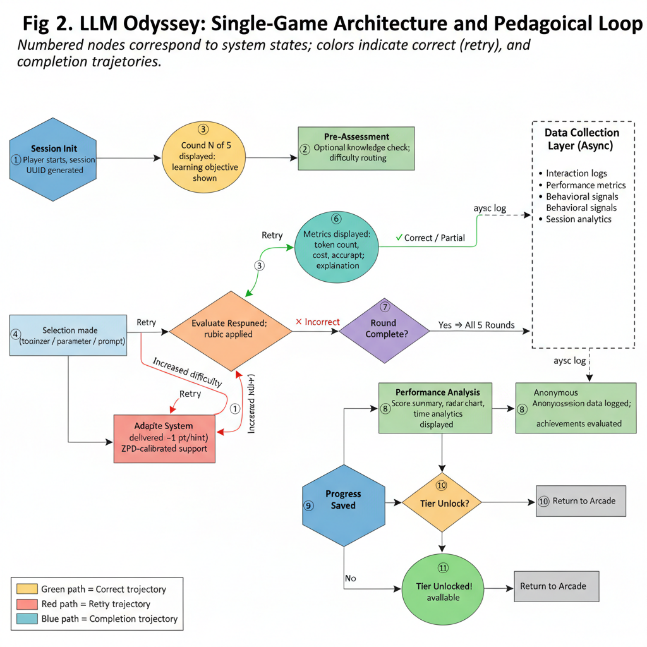}
\caption{LLM Odyssey: Single-Game Internal Architecture and Pedagogical Loop. Numbered nodes depict system states that facilitate progressive difficulty and scaffolded intervention. Color-coded trajectories (green: correct, red: retry, blue: completion) provide immediate formative feedback, guiding learners through authentic LLM engineering challenges and managing cognitive load.}
\label{fig:gameloop}
\end{figure}

Each game implements following five pedagogical strategies that are grounded in a distinct body of learning science literature and are expressed through specific mechanical choices in the game design (see Fig.~\ref{fig:gameloop} for detailed mapping of the game mechanics):

\begin{itemize}
\item \textit{Immediate Formative Feedback}: Each student action produces quantitative and qualitative output. Token counts, attention patterns, loss curves, and cost metrics update without delay. This design follows Black and Wiliam's \cite{black1998} evidence that immediate feedback during practice supports learning.
\item \textit{Scaffolded Hints}: An adaptive hint system provides graduated support, progressing from conceptual cues to partial solutions. Students incur a one point reduction per hint to encourage independent attempts first. This follows Vygotsky's principle \cite{vygotsky1978} that support should be calibrated to the learner's current capability. Whether the current calibration achieves this for diverse learners is a question for future evaluation.
\item \textit{Progressive Difficulty}: Each game comprises five rounds with increasing complexity. This draws on flow theory \cite{csikszentmihalyi1990}, which identifies an optimal zone between boredom and anxiety as the condition for sustained engagement. Whether the current difficulty curve achieves this in practice is an empirical question.
\item \textit{Worked Examples}: Integrated concept guides include annotated code, visual diagrams, and step by step explanations. Cognitive load theory \cite{sweller2011} predicts that well structured worked examples reduce extraneous load for novice learners, supporting schema acquisition.
\item \textit{Authentic Context}: Challenges are framed as production scenarios, for example: ``Your inference API costs \$50,000 per month. Reduce costs by 40\% while maintaining P95 latency below 200ms.'' Brown, Collins, and Duguid \cite{brown1989} argue that situating learning in authentic contexts supports transfer to real world tasks.
\end{itemize}

\subsection{Sample Game: Token Forge}
\begin{figure}[!t]
\centering
\includegraphics[width=0.9\columnwidth]{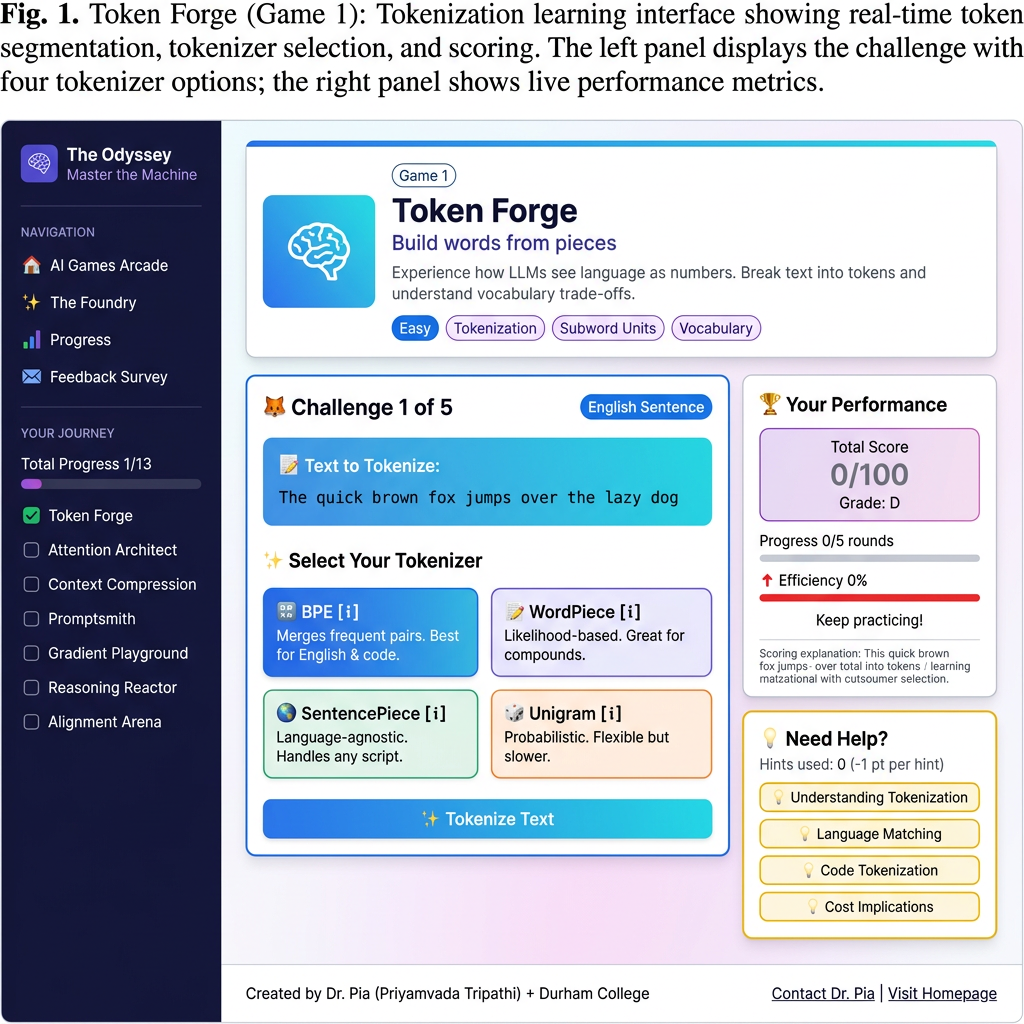}
\caption{Token Forge Interface.}
\label{fig:tokenforge}
\end{figure}

To concretely illustrate the platform's architectural choices and integrated pedagogical strategies, this section provides a detailed walkthrough of a representative game called Token Forge (shown in Fig.~\ref{fig:tokenforge}). By examining Token Forge, readers can observe how foundational concepts are introduced through interactive mechanics, immediate feedback, and progressive challenges.

Students receive text samples spanning multilingual content, Python code, and legal documents, and select among four tokenizers: BPE \cite{sennrich2016}, WordPiece \cite{schuster2012}, SentencePiece \cite{kudo2018sp}, and Unigram \cite{kudo2018sr}. Immediate feedback is instantiated through a color coded token segmentation display, a token count, a vocabulary efficiency score, and an estimated API cost that update on every selection. After each choice, students receive specific comparative feedback such as: ``SentencePiece produced 19 tokens versus 23 for WordPiece, a 17\% reduction; subword strategies may better capture code syntax.'' The hint system offers cues such as ``Consider which algorithm handles morphologically rich languages more efficiently'' before revealing algorithm comparisons. Five rounds increase in difficulty from simple English text to multilingual legal documents. An integrated concept guide connects tokenization algorithm theory to practical cost implications, and all challenge texts are drawn from realistic production scenarios. This same pattern of five interlocking pedagogical strategies, adapted to each topic's mechanics, is repeated across all 13 games.

\section{Implementation and Evaluation}
The platform uses a client side architecture to maximize accessibility. The frontend is built with React 18.2 and Tailwind CSS; game logic executes on the client, eliminating server round trips during gameplay and maintaining interaction response times below 16ms in testing. A serverless backend handles data persistence and analytics collection, with all records keyed to anonymous session identifiers via a localStorage UUID. No personally identifiable information is collected, supporting research ethics compliance. Measured deployment performance includes page load below 100ms and animations at 60 frames per second.

\begin{figure}[!t]
\centering
\includegraphics[width=0.9\columnwidth]{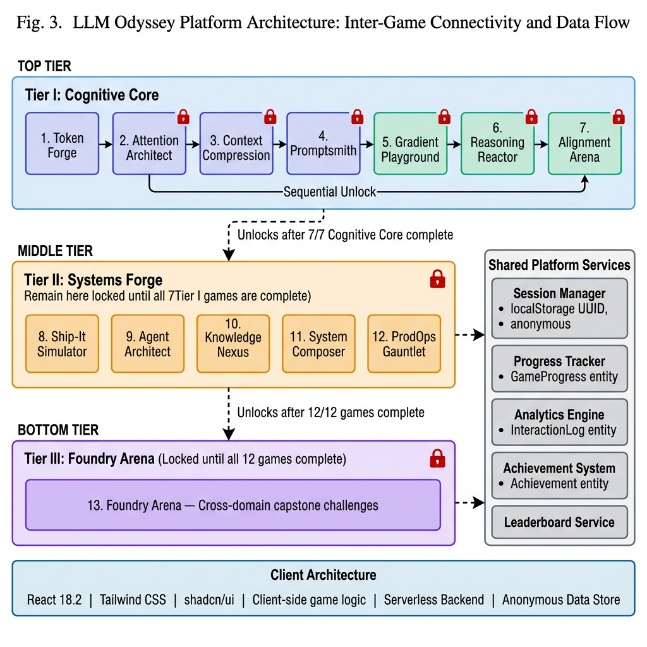}
\caption{LLM Odyssey Platform Architecture: Inter-game Connectivity and Data Flow.}
\label{fig:architecture}
\end{figure}

The platform logs anonymous interaction data to support future evaluation studies and potential adaptive features. Engagement data includes time on task per round, completion rates, game sequence patterns, and session duration. Performance data includes round scores, accuracy trends, and improvement across retry attempts. Behavioral data includes hint requests, error patterns, and retry frequency. Assessment data includes pre and post knowledge test scores across five LLM concept domains, and survey data captures validated Likert scale responses on perceived learning and difficulty. All records are keyed to anonymous session identifiers with no linkage to personally identifiable information, supporting research ethics compliance. Fig.~\ref{fig:architecture} details inter game connectivity and data flow for the three tiers.

\subsection{Design Challenges}
Three primary challenges shaped the platform's development.

The first challenge was balancing educational fidelity with computational feasibility. Full BPE tokenization is computationally intensive for browser based execution. The current implementation uses precomputed token breakdowns for challenge texts, preserving educational accuracy while maintaining performance; this trade off is presented to students as an example of a real engineering constraint.

The second challenge was difficulty calibration. Expert review and pilot observation indicated that a uniform difficulty curve is insufficient: novice students may become blocked on later rounds while experienced students may find early rounds insufficiently challenging. Difficulty curves have been adjusted based on these observations; an adaptive preassessment routing mechanism has been identified as the priority for a subsequent iteration.

The third challenge was production simulation authenticity. Systems Forge games must convey production engineering constraints without access to live infrastructure. The current approach uses parameterized scenarios with explicit numerical constraints, for example: ``Your API serves 10,000 requests per day. P95 latency budget: 200ms. Monthly cost ceiling: \$8,000.'' This is intended to provide an authentic decision context within browser based execution constraints.

\subsection{Preliminary Findings and Evaluation Protocol}
Following deployment at Durham College (Oshawa, ON, Canada) in Winter 2026, a structured usability and content review was conducted with two faculty members with computer engineering experience (N=2). This review was not a formal evaluation and does not constitute evidence of learning effectiveness. The reviewers identified the game based format and immediate quantitative feedback as strengths, noting that production scenarios in Systems Forge were relevant to current industry practice. The primary enhancement priority identified was adaptive difficulty: the reviewers noted that expertise variation within student cohorts is substantial, and that a preassessment routing mechanism would better serve students at both ends of the distribution.

Technical performance targets were met across four weeks of deployment, with page load consistently below 100ms and no availability incidents. This review provides initial evidence of technical feasibility and face validity of the pedagogical design; validation of learning outcomes requires a controlled study following the protocol below.

A formal mixed methods evaluation protocol has been designed for the platform. Owing to the author's transition from Durham College to Tufts University, execution of this study is not currently scheduled; the protocol is documented here so that it can be carried out, subject to institutional ethics approval, by the author or by institutions adopting the publicly available platform. The protocol specifies N=50 computer science students in an introductory AI course across 12 weeks, with 6 to 8 hours of self paced platform use over four weeks. Four data sources would be collected: (1) pre and post knowledge tests comprising 10 items across five LLM concept domains, with content validity reviewed by domain experts prior to administration; (2) automated engagement logs including completion rates, time on task, hint usage, and retry patterns; (3) a 44-item post platform survey using Likert scales and open ended items; and (4) semi structured interviews with 10 to 15 volunteers (30 to 45 minutes each).

The four research questions are:
\begin{itemize}
\item \textbf{R1}: To what extent does engagement with the LLM Odyssey platform lead to statistically significant improvements in students' pre- to post-assessment knowledge test scores concerning LLM engineering concepts?
\item \textbf{R2}: What distinct engagement patterns (e.g., time on task, game completion rates, use of pedagogical features) characterize student interaction with the LLM Odyssey platform?
\item \textbf{R3}: Which specific pedagogical elements integrated into the LLM Odyssey platform (e.g., immediate feedback, scaffolded hints, progressive difficulty) are significantly associated with enhanced learning outcomes and student engagement?
\item \textbf{R4}: How do students' learning and engagement trajectories within the LLM Odyssey platform vary as a function of their prior levels of expertise in Large Language Model engineering?
\end{itemize}

Quantitative data would be analyzed using a paired-samples t-test with Cohen's d for R1 to assess knowledge gains. R2, examining engagement patterns, would employ descriptive statistics and correlational analyses. Multiple regression would be utilized for R3 to identify pedagogical elements associated with outcomes, while R4 would be addressed through pre-test tertile subgroup analysis to explore trajectory differences based on prior expertise. Qualitative data would undergo thematic analysis by two independent coders, targeting an inter-rater reliability of Cohen's kappa $\geq$ 0.80. A primary limitation is the single-institution sample without a comparison group, which restricts causal inference; a comparative design would be a natural extension in subsequent studies.

\section{Conclusion}
This paper describes LLM Odyssey, an open source, browser based game based learning platform comprising 13 interactive games for LLM engineering education. The design addresses a documented gap in instructional resources for topics including tokenization, transformer architectures, prompt engineering, RAG, and production deployment. Platform design is grounded in constructivism, scaffolding theory, mastery learning, and cognitive load theory, with each strategy mapped to specific game mechanics. Initial expert review provides evidence of technical feasibility and face validity of the pedagogical approach; evidence of learning effectiveness has not yet been established and is the focus of the evaluation protocol presented here.

The designed evaluation protocol (N=50) specifies how initial data on pre to post knowledge gains, engagement patterns, and the association between specific pedagogical elements and learning outcomes can be collected. Causal claims about effectiveness will require studies across multiple institutions and cohorts. Potential enhancements include preassessment driven adaptive difficulty routing, LMS integration via SCORM 2004 and LTI grade pass back, and expansion to multimodal and reinforcement learning content. The platform is publicly available at \url{https://llmodyssey.piatripathi.ca} for community use and adaptation. Beyond the platform itself, this work offers the computing education community a transferable design pattern: tiered progression mapped to Bloom's revised taxonomy, with each game mechanic explicitly grounded in learning science, provides a replicable template for building instructional tools in other rapidly evolving technical domains where curricula lag industry practice.

\section*{Acknowledgment}
This work was conducted while the author was with Durham College, Oshawa, ON, Canada. The author is now with the Tufts Institute for Artificial Intelligence, Tufts University, Medford, MA, USA.

\end{document}